\def\a{\alpha}
\def\b{\beta}

\def\finalcopy{\pagestyle{empty}}

\def\@oddhead{}\def\@evenhead{}
\def\@oddfoot{\rm\rightmark \hfil Page \thepage}
\def\@evenfoot{\@oddfoot}

\def\onehead#1{\vskip\baselineskip\centerline{\normalsize #1}
               \vskip8pt}

\font\tencp=cmcsc10
\newfam\cpfam
\catcode`@=11
\newdimen\h@big
\newdimen\h@Big
\newdimen\h@bigg
\newdimen\h@Bigg
\newcount\f@ntkey            \f@ntkey=0
\def\caps{\fam\cpfam \tencp \f@ntkey=8 }
\def\samef@nt{\relax \ifcase\f@ntkey \rm \or\oldstyle \or\or
         \or\it \or\sl \or\bf \or\tt \fi}   
\font\tenrm=cmr10
\font\teni=cmti10
\font\tensy=cmsy10
\font\tenex=cmex10
\font\tensl=cmsl10
\font\tenbf=cmbx10
\font\tentt=cmtt10
\font\tencp=cmcsc10
\font\sevenrm=cmr7
\font\seveni=cmti7
\font\sevensy=cmsy7
\font\sevenbf=cmbx7
\font\fiverm=cmr5
\font\fivesy=cmsy5
\font\fivebf=cmbx5
\def\tenpoint{\relax
    \textfont0=\tenrm          \scriptfont0=\sevenrm
    \scriptscriptfont0=\fiverm
    \def\rm{\fam0 \tenrm \f@ntkey=0 }\relax
    \textfont1=\teni           \scriptfont1=\seveni
    \def\oldstyle{\fam1 \teni \f@ntkey=1 }\relax
    \textfont2=\tensy          \scriptfont2=\sevensy
    \scriptscriptfont2=\fivesy
    \textfont3=\tenex          \scriptfont3=\tenex
    \scriptscriptfont3=\tenex
    \def\it{\fam\itfam \teni \f@ntkey=4 }\textfont\itfam=\teni
    \def\sl{\fam\slfam \tensl \f@ntkey=5 }\textfont\slfam=\tensl
    \def\bf{\fam\bffam \tenbf \f@ntkey=6 }\textfont\bffam=\tenbf
    \scriptfont\bffam=\sevenbf     \scriptscriptfont\bffam=\fivebf
    \def\tt{\fam\ttfam \tentt \f@ntkey=7 }\textfont\ttfam=\tentt
    \def\caps{\fam\cpfam \tencp \f@ntkey=8 }\textfont\cpfam=\tencp
    \setbox\strutbox=\hbox{\vrule height 8.5pt depth 3.5pt width\z@}
    \samef@nt}
\documentstyle[twocolumn,12pt]{article}
\def\@maketitle{\vbox
to \titleblockheight{
\hsize\textwidth
  \linewidth\hsize \vfil \centering
  {\large \@title \par} \vskip\baselineskip 
  {\normalsize\begin{tabular}[t]{c}\@author
  \end{tabular}\par}
  \vfill}}
\def\abstract#1{\vbox{
\twocolumn[\vbox
{\hsize\textwidth\parindent0pt\leftskip.75in\rightskip.75in
\vskip\baselineskip
\centerline{\normalsize Abstract}\vskip6pt
#1\par\vfil\vskip\baselineskip}
]
}}
\date{ }
\title{\normalsize
\begin{flushright}{UPR-524T}\end{flushright}
NON-PERTURBATIVE INTERACTIONS IN STRING THEORY
\thanks{Supported in part by the Department of Energy under  contract
No. DOE-AC02-76-ERO-3071.\hfill\break Talk presented by B. Ovrut
at the XXVI International Conference on High Energy Physics,
Aug. 1992, Dallas, Texas
.}}
\author{Ram Brustein and Burt A. Ovrut\\
        Department of Physics \\
        University of Pennsylvania\\
        Philadelphia, PA 19104}

\begin{document}
\finalcopy

\maketitle

\abstract{New non-perturbative interactions in the effective action of
 two dimensional  string theory are described.
These interactions are due to ``stringy" instantons.\hfill\break} 

\vskip-2pc
\onehead{ INTRODUCTION}
Non perturbative effects in string theory are important. To someone
familiar with  weakly  coupled quantum field theories that may sound  a little
strange. After all, non-perturbative effects are much smaller than
perturbative effects in such theories.
Typically, in a  weakly coupled field theory,
non-perturbative effects have strength  $e^{-{1/g^2}}$ compared to
perturbative quantities that have strength $g^2$, where $g$ is the coupling
parameter of the theory. For example if $g^2=.3$ the ratio of these numbers
is about $1:100$. The only case in which non-perturbative quantities are
important is when their value in perturbation theory vanishes. This is
the situation in superstring theory for a number of important quantities,
e.g., the size of supersymmetry  breaking$^{(1)}$.
The smallness of non-perturbative effects is actually an advantage in string
theory because it may explain  the small ratio of
the electroweak scale and Planck scale.
\vfill\eject
To confront string theory with the real observable world we have to
understand the source of non-perturbative interactions in string theory
 and know how to calculate them.
Matrix models, and especially $d=1$ matrix models$^{(2)}$,
 offer  a unique opportunity to obtain some  insight into non-perturbative
string theory. Certain matrix models have associated with them very
simple string theories with a low number of degrees of freedom, propagating
in a low number of space-time dimensions.
The $d=1$ matrix model is the most complicated matrix model which can still
be exactly solved.
On the other hand, it describes the simplest space-time dynamics which
is still interesting. In the double scaling limit, the $d=1$ matrix model
 describes strings propagating in one time dimension
and one spatial dimension. An equivalent description is given in terms
of a bosonic collective field theory$^{(3)}$ in $1+1$ dimensions of
one massless field. Notable features of collective field theory is that
the kinetic energy  is not canonical and the theory is not Lorentz invariant.

The $d=1$ matrix models, or the equivalent field theories
have the power to describe non-perturbative phenomena in the associated
$1+1$ string theories.  This is interesting by itself.
However, there may well be some general  features of non-perturbative string
theory that are common to all string theories, including
more complicated theories in higher dimensions such as $d=4$.
By studying the generic features of non-perturbative
behaviour in $1+1$ dimensional string theories, as we do in this talk,
one may learn about more realistic 4-dimensional string theories.
It is of interest to ask whether or not there is any indication in string
theory of common, non-perturbative behaviour. The answer$^{(4)}$ is yes!

First recall that in quantum field theory there is a well known connection
 between the  large order behaviour of amplitudes and non-perturbative
effects. Typically, amplitudes grow as   $G!$  where $G$ is the number
of loops, while non-perturbative effects
  have strength  $e^{-{1/g^2}}$, where $g$ is the coupling constant of the
theory.  Both of these facts follow from
 the existence of  non-trivial classical solutions  of the equations of motion
of the field theory  in Euclidean space,
i.e. instantons.
The magnitude of the non-perturbative
effects due to non-trivial solutions in a field theory with one dimensionless
coupling parameter $g$ can  be estimated using a simple scaling argument.
Since the coupling parameter  in this case  can be scaled away, the
action can be written as $S(\phi,g)={1/g^2} \widetilde S(\widetilde\phi)$,
where $\widetilde S$ does not depend on $g$.
Therefore, any classical Euclidean solution with finite action has an action
of order ${1/g^2}$.
The magnitude of large order terms
in the perturbative expansion can also  be estimated by counting Feynman
diagrams. The number $G!$ basically comes from the number of diagrams.

Large  order growth of perturbative
amplitudes is a common feature of matrix models
 and more complicated string theories$^{(4)}$.
For a review of large order behaviour of matrix model amplitudes see ref.(5).
All matrix models, as well as critical bosonic string theory in 26 dimensions,
exhibit a strange phenomenon. The magnitude of $G$'th order amplitudes in
perturbation theory  grow like $(2G)!$.
It turns out that, in much the same way  as $G!$ behaviour corresponds to
$e^{-{1/g^2}}$ non-perturbative effects in quantum field theory,
in matrix models the large order  $(2G)!$ behaviour,
would correspond to non-perturbative effects of strength $e^{-{1/g}}$.
How do these peculiar effects arise?
In matrix models, there is a new type of instanton, involving
a single eigenvalue, that is responsible for these effects.

We expect that string theory is described at low energies by an effective
field theory with one dimensionless coupling parameter.
In view of the above scaling argument in quantum field theory,
it is of interest to ask how an  instanton action of order ${1/g}$ can ever
arise,
in such an effective field theory.
We found  that in matrix models, the associated effective action
does not obey the same scaling argument,
$S(\phi,g) \ne {1/g^2} \widetilde S(\widetilde\phi)$. Instead,
one finds that $g$ cannot be completely scaled out of $\widetilde S$
due to ``scale breaking terms". That is
$S(\phi,g) = {1/g^2} \widetilde S(\widetilde\phi,g)$. It follows
that a non-trivial solution can be a function of $g$. Furthermore, if
for such a solution $\widetilde S\sim g$, then $S\sim {1/g}$.
This is exactly what happens for one eigenvalue instantons.

In this talk, the role of these ``stringy" instantons in
the effective field theory of string theory  is discussed and
the non-perturbative interactions that they produce are derived.
For a more detailed and extensive discussion of the issues in this talk,
as well as a comprehensive list of references,
 the interested reader is refered to the original papers$^{(6)}$.

\onehead{EFFECTIVE FIELD THEORY FOR STRINGS}
Collective field theory for $d=1$ matrix models is written in terms
of the  density of eigenvalues,
$\phi(x,t)=\sum\delta(x-\lambda_i(t))$ where $\lambda_i$
are the eigenvalues of the matrix. The size of the matrix, $N$, is very large.
The field $\phi$ is called the  collective field.
The  Lagrangian  density  of  this collective field theory  is
\begin{equation}
{\cal L}_{\rm eff}=
{1\over 2}{\int\limits^x\dot\phi \int\limits^x\dot\phi \over\phi}
-{\pi^2\over 6} \phi^3-{1\over 2} ({1\over g}-x^2)\phi
\end{equation}

It is obvious that the field $\phi$ does not have canonical kinetic energy
and that the theory is not Lorentz invariant.
These deficiencies will be corrected soon. For a review of perturbative
results of collective field theory see ref.(7).
The classical equations of motion are not the usual field equations$^{(6)}$.
Instead  they are
\begin{equation}
\partial_x\!\left(\!
\int\limits^x\! dy\partial_t{\int\limits^y\dot\phi\over\phi}
\!-\!{\int\limits^x\dot\phi \int\limits^x\dot\phi \over 2 \phi^2}
 \!-\!{\pi^2\over 2}\phi^2
\!-\!{1\over 2}({1\over g}\!-\!x^2)\right)_{\!\!|x=\lambda_i(t)}
\hspace{-2pc}=0
\end{equation}
where the index $i$ now runs over $i=1,...,N\rightarrow\infty$.

These equations allow solutions of high density regions where $\phi>>1$
and low density regions.
The static, high density, solution of these equations is very simple
\begin{equation}\phi_0= {1\over\pi}\sqrt{x^2-{1\over g}} \end{equation}
where $|x|\ge\sqrt{{1\over g}}$. This static solution is displayed in Figure 1.

\ \vspace{2pc}
%
\vspace{15pc}
%
\vbox{\parindent=0pt\baselineskip=12pt\tenpoint Figure~1. The potential and
static, high density,\break\hbox to 42pt{\hfil} solution of collective  field
theory.}

There are also interesting time dependent Euclidean solutions to Eq.(2).
\begin{equation}
\phi_{inst}(x,\theta)=\delta\left(x-{1\over\sqrt{g}}
cos(\theta-\theta_0)\right)
 \end{equation}
This is an instanton that corresponds to tunneling of one eigenvalue
 across the barrier from
$x=\sqrt{{1\over g}}$ at Euclidean time $\theta=\theta_0$ to
$ x= -\sqrt{{1\over g}}$ at $\theta=\theta_0+\pi$. Note that the classical
solution in Eq.(4) is not a solution of the Euclidean continuation  of
of the unconstrained  field theory equations of motion.
The action of this one eigenvalue instanton is $\pi/g$, as mentioned in the
introduction.

The collective field theory Lagrangian density, expression (2), has
two notable deficiencies. First, the kinetic energy term is not in canonical
form.
This means that we have not identified correctly the canonical field of the
theory.
Second, and more important, the  coordinate $x$ appears  in the
potential energy and therefore Lorentz invariance seems to be broken
explicitly.
We remove both deficiencies. The first, following  ref.${(3)}$,
by  a field  redefinition $\partial_x\zeta=\phi-\phi_0$ and a coordinate
redefinition $x\rightarrow\tau={1\over\pi}\int\limits^x {dy\over\phi_0}$.
The second, following ref.${(8)}$,
 by enlarging the theory to include a new field, $D$. The non-trivial
vacuum expectation value of this  new field is responsible for the spontaneous
breaking of  Lorentz invariance. The resulting action  for the fields
$D$,$\zeta$
 in the new coordinates is
\begin{eqnarray}
&{\cal S}&=  \int dt d\tau \Biggl\{  {1\over 2}
{\nabla \zeta\cdot\nabla \zeta\over 1+{ 2\sqrt{\pi} }
{ e^{D}\over\left(1- {1\over {g}} e^{D}\right)^2 }\nabla\zeta\!\cdot\!\nabla D}
\nonumber\\
&-&\!{\sqrt{\pi}\over 4 }{ e^{D}\over\left(1- {1\over {g}} e^{D} \right)^2 }
{(\nabla \zeta\cdot\nabla D)^3\over 1+{ 2 \sqrt{\pi} }
{ e^{D}\over\left(1-{1\over {g}} e^{D} \right)^2 }\nabla \zeta\cdot\nabla D}
\nonumber\\ & - &{ \sqrt{\pi}\over 12}
{e^{D}\over\left(1-{1\over {g}} e^{D} \right)^2 }
(\nabla \zeta\cdot\nabla D)^3 \\ &-&
{1\over  384 \pi} e^{-2 D}
 \left[1-{1\over {g}} e^{D} \right]^4
\left[ {(\nabla D)^2-4} \right]
\Biggr\} \nonumber
\end{eqnarray}
The kinetic energy for $\zeta$ is clearly canonical,
and Lorentz invariance is manifest. The action is also invariant under
an unexpected$^{(9)}$ shift symmetry $\zeta\rightarrow\zeta+const.$
The field $\zeta$ has no potential, only derivative interactions.
This is similar to the dilaton interactions in the critical bosonic string
and the  corresponding complex field, $S$, in the superstring. The parameter
$g$ appears in the action (5) only through the coupling parameter of the
effective theory, which is $\hbox{\it g} (D)=4\sqrt{\pi}
 { e^{D}\over\left(1- {1\over {g}} e^{D} \right)^2 }$. Thus $g$ cannot be
absorbed into a redefinition of $D$ and scaled away from the effective action.
The coupling parameter $\hbox{\it g}(D)$ is an example of the ``scale breaking"
terms that were discussed in the introduction.

The general solution of the equations of motion derived from (5) is given by
\begin{equation}
<D>  =   a (t-{\bar t}) + b (\tau -{\bar\tau}),\ \
<\zeta>  ={1\over g}+c \end{equation}
where $a,b,c,{\bar t}$ and ${\bar\tau}$ are real parameters, $b^2-a^2=4$ and
$c,{\bar t},{\bar\tau}$ are arbitrary.
An interesting vacuum solution, which is a combination of two solutions
is $<\zeta>={1\over g}$ and  $<D>=-2\tau$ for $\tau\ge\ln\sqrt{{1\over g}}$,
henceforth called region $I$, and $<D>=-2(\tau-2 \ln\sqrt{{1\over g}}+\pi)$
for $\tau\le\ln\sqrt{{1\over g}}-\pi$, henceforth called
region $III$. The effective coupling parameter in region $I$ is
${\hbox{\it g}}_-$
 and in region $III$ is ${\hbox{\it g}}_+$.
The spatial interval  $\ln\sqrt{{1\over g}}-\pi\le \tau\le \ln\sqrt{{1\over
g}}$
is called region $II$. We plot the effective coupling parameters in Figure 2.

\ \vspace{1pc}
\vspace{15pc}
\vbox{\parindent=0pt\baselineskip=12pt\tenpoint Figure~2.
Effective coupling parameters in different\break\hbox to 42pt{\hfil}
 regions of space.}
%
The two dashed ``walls" in Figure 2 mark the boundaries of the regions.
As can be seen from Figure 2,  in the regions far away from the two walls,
the coupling parameters are small and we expect the effective field theory (5)
to provide a good description of physics. Near the ``walls" the coupling
parameters blow up. Region $II$ is unknown territory.

We use low density collective field theory as a guide in the unknown
territory, region $II$. Comparing the low density collective field theory
solution to the solution of the effective field theory we see that in
region $II$ their solutions should be the same.

\ \vspace{2pc}
\vspace{20pc}
\vbox{\parindent=0pt\baselineskip=12pt\tenpoint Figure~3.
Instanton tunneling between regions $I$ \break\hbox to 42pt{\hfil}
and $III$. The static solutions  are also shown.}

In Euclidean space the solution is therefore  the
instanton of Eq.(4),  expressed in the new coordinates
\begin{equation}\hspace{-1pc}\phi_{inst}(\tau,\theta)=
{1\over\sin(\theta-\theta_0) }
\delta\!\left(\!\tau\!-\![\ln\sqrt{{1\over g}}\!-\!(\theta\!-\!\theta_0)]
\right) \end{equation}
This is an instanton which corresponds to the tunneling of a single
eigenvalue across the barrier from $\tau=\ln\sqrt{ {1\over g}}$ at
$\theta=\theta_0$ to $\tau=\ln\sqrt{ {1\over g}}-\pi$ at $\theta=\theta_0+\pi$.
 Note that the velocity of the eigenvalue
at either side of the barrier vanishes. Therefore,
the Euclidean conjugate momentum of the
instanton in region $II$, matches continuously at the boundaries with the
vanishing  conjugate momentum of the static  vacua $\phi_0$
in regions $I$ and $III$.
We represent this tunneling process in Figure 3.

The action (5) is the effective space-time action that corresponds to
string theories described by the following $\sigma$-model$^{(7,8)}$
\begin{eqnarray}
&I={1\over 4\pi}\int d^2z\sqrt{\hat g} \biggl\{ &\hat g^{\a\b}G_{\mu\nu}
\partial_{\alpha}X^{\mu}\partial_{\beta}X^{\nu}+ \nonumber \\
&\ \ \ \ \   &\hat R D(X)+2 T(X)\biggr\} \end{eqnarray}
where $\hat g_{\a\b}$ is the fixed world sheet metric  with Euclidean
signature and  $\hat R$ is the corresponding Ricci scalar.
The sigma model  field $X_\mu$  stands for  two  scalar
fields, $X_0 (z)$, and $X_1(z)$.
The   field $G_{\mu\nu}(X)$ is the target space metric,  assumed here  to  have
Euclidean signature, $D(X)$ is the dilaton,  and $T(X)$ is the tachyon.
The field $\zeta$ is related to the tachyon $\zeta\propto T e^{-D}$.
The instanton shown in Figure 3 is therefore a ``stringy instanton".

\onehead{NON-PERTURBATIVE INTERACTIONS}

We  integrate over the instantons and represent
their effects as effective terms in the $D$,$\zeta$ theory. Since $\zeta$ is
the light field we restrict our attention to $\zeta$ operators.
The effective operators are especially important. They provide the
bridge between the discrete sector of the theory and the continuous sector.
A more detailed analysis is given in  ref.$(6)$.

The  instanton has three parameters, $\bar\tau$, $\theta_0$ and $\alpha$.
The parameters $\bar\tau$ and $\theta_0$ were  defined in Eqs.(6) and (7).
The parameter $\alpha$ is  related to the  parameter
$a$ in the Euclidean space continuation of Eq.(6),   $a=2\sin\alpha$.
Changing $\alpha$ results in the rotation of the vacuum solution in
$\tau-\theta$
space. There are three zero modes corresponding to  the three broken generators
of the Euclidean group associated with $\bar\tau$, $\theta_0$, $\alpha$.
These have to be integrated and produce a volume
factor $Vol\propto \int d\bar\tau d\theta_0  d\alpha$.

The dilute gas summation
over instantons induces  effective terms in the $D$,$\zeta$ Lagrangian.
The most general  action   induced by instantons is
$\Delta S=\int d\tau d\theta  \{\sum\limits_n C_n O_n(\tau,\theta)\}$,
where $O_n$ are local operators built from  $D$ and $\zeta$ and
their derivatives.
The coefficients $C_n$ can be computed by expanding the action around the
instanton background.
All the coefficients $C_n$ are proportional to the universal factor of the
exponent of the instanton action  and the remaining
factor depends on the  particular operator that is considered.
 Since the ``size" of the instanton is $\sqrt{g}$,
the dimension of the operator determines the $ g$ dependence of $C_n$. That is
\begin{equation}C_n= \hbox{\it \~C}_n g^{d(n)} e^{-{\pi\over g} }\end{equation}
where $d(n)=[{\rm dimension}(O_n)]^{1\over 2}-1$
and $\hbox{\it \~C}_n$ is a numerical coefficient.  The coefficient
$\hbox{\it \~C}_n$ is not expected to be particularly large or particularly
small.

We are interested in large ${1\over g}$ that corresponds to small $g$.
In that case the dominant and most interesting operator  is the unit operator.
All other operators are suppressed by powers of $ g$.
The coefficient  of the unit operator is given by
$C_0= \hbox{\it \~C}_0 {\scriptstyle {1\over g}} e^{-{\pi\over g} }$.
This result  was obtained in the background of a constant
field $<\zeta>={1\over g}$. Lorentz
invariance then dictates that at least for slowly varying fields the effective
operator depends on the full field $\zeta$ and not just its constant mode
${1\over g}$.  Therefore the final result of the  induced operator is
\begin{equation} \Delta{\cal L}_0=\hbox{\it \~C}_0\zeta e^{-\pi\zeta}
\end{equation}
written in terms of the tachyon $T$ the induced operator is
\begin{equation} \Delta{\cal L}_0=\hbox{\it \~C}_0 Te^{-D} e^{-\pi T e^{-D}}
\end{equation}

This operator breaks the  $\zeta$ shift symmetry. It induces a  runaway
non-perturbative potential for the field $\zeta$. Similar effects due to
field theoretic non-perturbative interactions occur in more complicated string
theories$^{(1)}$. Recall that in more complicated theories the dilaton
has only derivative interactions and no potential.
Known field theoretic non-perturbative interactions induce a runaway
potential for  the dilaton and break the dilaton shift symmetry.
These effects, however,
are of typical field theory strength i.e. $e^{1/g^2}$. In the superstring,
the appearance of a potential for the $S$ field signals supersymmetry breaking.

It is therefore tempting to conjecture that stringy instantons similar to our
stringy instantons appear in 4-dimensional superstring theories and that they
induce non-perturbative operators of the type discussed above.
In that case these operators are expected to be proportional to
the universal factor $e^{-\sqrt{S}}$,  where $S$ is a complex field
that naturally appears in the effective low energy supergravity  field
theory obtained from superstring theory. The dilaton is related to
the real part of $S$, $<Re S>\sim{1\over g^2}$.
Note that the non-perturbative effects considered previously in the
literature induced operators of field theory strength, and therefore
are proportional to the  universal factor  $e^{-S}$.
Since the coupling parameter $g$ is expected to be small, the difference
between these two universal factors can be big.
If indeed such non-perturbative `stringy" interactions occur, they  may have
important phenomenological consequences.
%

%



\end{document}